\documentclass{article}

\usepackage[nonatbib,preprint]{neurips_2025}

\usepackage[utf8]{inputenc} % allow utf-8 input
\usepackage[T1]{fontenc}    % use 8-bit T1 fonts
\usepackage{hyperref}       % hyperlinks
\usepackage{url}            % simple URL typesetting
\usepackage{booktabs}       % professional-quality tables
\usepackage{amsfonts}       % blackboard math symbols
\usepackage{nicefrac}       % compact symbols for 1/2, etc.
\usepackage{microtype}      % microtypography
\usepackage{xcolor}         % colors
\usepackage[numbers]{natbib}
\usepackage{amssymb}
\usepackage{amsmath}
\usepackage{enumitem}
\usepackage[ruled, lined, linesnumbered, commentsnumbered, longend]{algorithm2e}
\usepackage{graphicx}

\title{The End Of Universal Lifelong Identifiers: Identity Systems For The AI Era}

\author{%
  Shriphani Palakodety \\
  Onai Inc.\\
  San Jose, CA 95129 \\
  \texttt{spalakod@onai.com} \\
}

\begin{document}

\maketitle

\begin{abstract}
Many identity systems assign a single, static identifier to an individual for life, reused across domains like healthcare, finance, and education. These Universal Lifelong Identifiers (ULIs) underpin critical workflows but now pose systemic privacy risks. We take the position that ULIs are fundamentally incompatible with the AI era and must be phased out. We articulate a threat model grounded in modern AI capabilities and show that traditional safeguards such as redaction, consent, and access controls are no longer sufficient. We define core properties for identity systems in the AI era and present a cryptographic framework that satisfies them while retaining compatibility with existing identifier workflows. Our design preserves institutional workflows, supports essential functions such as auditability and delegation, and offers a practical migration path beyond ULIs.
\end{abstract}

\section{Introduction}

Universal Lifelong Identifiers (ULIs) are persistent identifiers assigned once and reused across domains without scoping or expiration. Examples include the Social Security Number (SSN)~\cite{ssb_ssn_history} and Aadhaar~\cite{aadhaar}. ULIs underpin services across healthcare, finance, education, and law enforcement. Similar identifiers are issued by large digital platforms to support personalization, access control, and tracking. Reused without isolation, ULIs function as universal join keys and enable linkage across datasets, institutions, and time.

We take the position that \textbf{Universal Lifelong Identifiers (ULIs) pose unprecedented privacy risks in the age of advanced AI systems and must be systematically phased out}. The persistent, cross-domain use of ULIs has long raised privacy concerns and prompted legislation~\cite{ssb_ssn_history,hipaa1996}. With modern machine learning systems, particularly large language models (LLMs) and sophisticated computer vision tools, traditional safeguards are no longer sufficient. The accelerating capabilities of these AI models in extracting identifiers from unstructured data~\cite{carlini_extracting}, linking records across contexts using learned representations~\cite{narayanan2008robust}, and demonstrably memorizing sensitive identifiers during their training phase pose systemic risks that can be exploited even by actors with limited resources. Existing mitigation techniques~\cite{machine-unlearning-survey,learning-to-unlearn,locate-edit-gpt-fact} are approximate and unauditable in practice.

ULIs routinely appear in plaintext across physical forms (which are frequently digitized), logs, and a multitude of digital documents and data interchange files due to regulatory and operational requirements~\cite{purpose_of_ssn,unique_health_id}. These patterns reflect institutional convenience and were once acceptable under the assumption that identifiers could be scoped to specific contexts, redacted if leaked, or remedied through legal recourse~\cite{solove2006taxonomy,ohm2010broken}. Today, they form an attack surface that modern AI systems are uniquely equipped to exploit.

ULIs often enter circulation through data breaches, scraped documents, or leaked forms, many of which are later posted publicly or traded on dark web marketplaces~\cite{ssn-capital-one,srinivasan2019equifax}. AI tools such as OCR engines and document parsers can extract identifiers from unstructured data~\cite{edpb_ocr_2024,tr-ocr}, feeding them into large-scale model training corpora. The opacity of large-scale training pipelines, which often use proprietary datasets and undisclosed preprocessing steps, makes it difficult to audit whether ULIs have been ingested. Their complete and verifiable removal remains an open research problem~\cite{preventing-false-sense-privacy} and typically requires costly, impractical model retraining. Even partial or redacted identifiers can be linked through AI-based inference~\cite{ohm2010broken,reidentification-risk}. Breaches that were once localized now become permanent features of deployed models. These risks cannot be mitigated through downstream fixes or post-hoc filtering and require a rethinking of identity systems.

Our main contributions, detailed in the following sections, are:

\begin{itemize}
  
  \item We define an AI-centric threat model for ULIs, demonstrating how modern machine learning capabilities, such as large-scale data ingestion, model memorization, and AI-driven linkage, render traditional safeguards insufficient.
  
  \item We derive essential properties for privacy-preserving identifiers suited for the AI era that serve as a drop-in replacement for ULIs.
  
  \item We present a conceptual cryptographic framework that satisfies these properties, offering stronger individual privacy while reducing the exposure of sensitive identifiers to large-scale AI systems.
  
\end{itemize}

\section{Background and related work}

\textbf{Legal and philosophical foundations}: Legal scholarship establishes that robust identity protections are necessary for individual autonomy~\cite{brandeis_warren,cohen_configuring,regan_legislating_privacy}. The contextual integrity framework~\cite{nissenbaum} formalizes how information flows should respect contextual norms and expectations, while privacy taxonomies~\cite{solove2006taxonomy} categorize harms including aggregation and secondary use. An analysis of ULIs through these lenses reveals that they violate contextual boundaries and individual privacy.

\noindent \textbf{AI capabilities and resulting threats}: Recent work demonstrates that AI systems can extract and memorize personal identifiers from training data~\cite{carlini_extracting,lukas-pii-llm,quantifying_memorization}. Large language models have been shown to reproduce specific identifiers when prompted~\cite{panda2024teach}, with real-world incidents confirming such exposures in deployed systems~\cite{livemint2023chatgptleak}. Advances in OCR now enable automated extraction of identifiers from handwritten forms and unstructured documents at scale~\cite{edpb_ocr_2024,tr-ocr}, possibly feeding these identifiers into training corpora. The closed nature of training pipelines makes PII filtering unverifiable, even with specialized probing tools~\cite{carlini_extracting,pro-pile}.

\noindent \textbf{Privacy risks of persistent identifiers}: Prior work has established important principles for identifier privacy, showing how persistent identifiers enable cross-service tracking~\cite{nikiforakis2013cookieless,eckersley2010panopticlick,reardon2019fifty} and reliable re-identification across datasets~\cite{narayanan2008robust,sweeney2000simple,ohm2010broken}. ULIs extend these concerns to mandatory institutional systems where legal requirements across critical services—healthcare, finance, employment—create systemic vulnerabilities that individuals cannot avoid~\cite{gao_ssn_laws}. Our work addresses the architectural problem of persistent identifiers in mandatory institutional systems.

\noindent \textbf{Anonymous credentials and cryptographic identity systems}: Anonymous credential systems enable selective attribute disclosure without revealing identity~\cite{camenisch2001efficient,uprove}. They have been successfully deployed in messaging applications~\cite{signal_auth} and government identity systems~\cite{alpar2017irma}. Our approach can be considered an anonymous identifier generation scheme that adapts these principles while acknowledging institutional inertia that requires ULIs to be entered in forms, printed on documents, and shared with organizations~\cite{eid-poor-adoption}. We maintain workflow compatibility while providing cryptographic protection against extraction and correlation.

\noindent \textbf{Self-sovereign identity.} Government identity systems have increasingly adopted self-sovereign identity principles, which emphasize user control over digital credentials~\cite{allen_ssi}. Recent deployments~\cite{alpar2017irma,german-id} demonstrate institutional interest in cryptographic identity systems that support user custody. Our approach applies self-custody principles to replace ULIs.

\noindent \textbf{Usability and adoption of privacy technologies}: Research on privacy technology adoption shows that workflow disruption is a primary barrier to deployment~\cite{chi-examine-abandon-security,eval-usability-privacy-choice,eidas-challenges}. This informs our approach of designing cryptographic protections around existing identifier workflows.

\noindent \textbf{PII mitigation in AI systems}: A growing body of work addresses the presence of PII in machine learning pipelines. Preprocessing tools aim to detect and redact sensitive fields from unstructured data before training~\cite{scrubadub,nvidia_nemo,hydroxai_pii_masker}. Post-hoc approaches include machine unlearning~\cite{machine-unlearning-survey,learning-to-unlearn}, and model editing techniques~\cite{locate-edit-gpt-fact,meng2023memit}. Alignment methods have also been applied to discourage models from disclosing PII in responses~\cite{ouyang2022training,privacy-preserving-instructions}. However, these mitigations remain approximate, expensive, and difficult for end users to verify or audit. Our proposal instead advocates for identity systems that are inherently robust to evolving AI capabilities.

\section{Threat model: AI risks to ULIs}

\subsection{Initial assumptions} 
The design and deployment of ULIs historically rested on five key assumptions:

\begin{itemize}[leftmargin=*]
    \item \textbf{Limited scope}: Identifiers would be used within specific domains rather than reused universally~\cite{gao_ssn_laws}.
    
    \item \textbf{Trusted custodians}: Identifiers would be shared primarily with high-trust institutions such as government agencies, banks, and healthcare providers~\cite{gao_ssn_laws}. Such institutions were believed to have rigorous privacy and security practices.
    
    \item \textbf{Controlled linking}: Cross-domain usage would require explicit user consent or formal judicial authorization~\cite{justice_privacy_act}.
    
    \item \textbf{Limited adversaries}: Non-state actors would face significant barriers to extracting or correlating identifiers across unstructured data at scale~\cite{bigdata_threat_model}.
    
    \item \textbf{Manageable breaches}: Identifier leaks would be detectable and traceable, with clear remediation paths through regulatory enforcement or technical countermeasures~\cite{ico_breach_response}.
\end{itemize}

\subsection{Breakdown of assumptions in the AI era}
In the age of large-scale AI and ML systems, these foundational assumptions have collapsed in the following ways:

\begin{itemize}[leftmargin=*]
    \item \textbf{Systemic proliferation}: Modern institutions are legally mandated to collect and process ULIs across domains. Regulations require these identifiers for employment, taxation, healthcare, and financial services~\cite{gao_ssn_laws,purpose_of_ssn,sin_employers,aadhaar}. Even privacy regulations~\cite{hipaa1996} contain broad exemptions that preserve these requirements. This legally-enforced ubiquity transforms static identifiers into de facto \textit{universal join keys} that individuals cannot avoid generating, creating permanent, cross-domain vulnerabilities. The consistency and frequency of identifier reuse produces a dense, high-quality signal that modern AI systems can ingest, memorize, and correlate across contexts.

    \item \textbf{Custodial concentration}: Structured PII is now concentrated in a small number of high-trust institutions, like governments, financial services, healthcare systems, creating systemic risk. These custodians were assumed to maintain strict controls, but repeated breaches~\cite{ssn-capital-one,srinivasan2019equifax} show that even regulated entities are frequently compromised. A single breach can release millions of clean, linkable identifiers, primed for AI-based extraction and misuse.

    \item \textbf{Unbounded adversary capabilities}: Foundation models now extract and correlate identifiers across document types with minimal effort~\cite{liang2023holistic}. Advanced OCR processes handwritten forms~\cite{edpb_ocr_2024,tr-ocr,htr-vt}, multimodal systems analyze mixed text and images~\cite{lee2024vhelm}, and specialized models extract structured data from tables~\cite{ashurytahan2025mightytorrbenchmarktable}. These commoditized tools allow even low-resourced actors to process vast document collections and perform cross-context linking that once required institutional expertise~\cite{staab24beyond}.
    
    \item \textbf{Irreversible exposure}: Large language models demonstrably memorize and can be prompted to regurgitate personal identifiers from their training data~\cite{carlini_extracting,carlini2019secret,lukas-pii-llm}. Recent studies quantify this memorization at the entity level, showing models retain specific identifiers with high fidelity~\cite{quantifying_memorization,pro-pile,Smith2023IdentifyingAM}. This creates a fundamentally new breach category: silent, permanent, and jurisdictionally unbounded. Closed training pipelines obscure the extent of PII preprocessing, making the scale of exposure unknowable even for open-source models. Alignment techniques intended to prevent PII disclosure~\cite{privacy-preserving-instructions} have been repeatedly circumvented through adversarial prompting~\cite{panda2024teach,preventing-false-sense-privacy}, with real-world incidents confirming retrieval of personal data from commercial systems~\cite{livemint2023chatgptleak}. Unlike traditional breaches, this exposure cannot be remediated through regulatory action or content removal once the model is distributed.

    \item \textbf{Accelerating capabilities}: The rapid trajectory of AI development~\cite{time2023aiprogress,liang2023holistic} points to models with increasingly powerful recall and reasoning capabilities. This trend suggests future systems will only enhance the extraction, memorization, and exploitation of personal identifiers.
\end{itemize}

\subsection{Adversary classification}
We identify four adversary classes with increasing capabilities and resources:

\begin{itemize}[leftmargin=*]
    \item \textbf{Commodity AI users:} Can access public LLMs and basic extraction tools to retrieve memorized identifiers from models~\cite{carlini_extracting,panda2024teach}.
    
    \item \textbf{Breach aggregators:} Combine leaked datasets with AI tools to build comprehensive profiles across contexts~\cite{reidentification-risk,narayanan2008robust,sweeney2000simple}.
    
    \item \textbf{Model trainers:} Inadvertently memorize ULIs during training, creating persistent exposure through model weights~\cite{quantifying_memorization}.
    
    \item \textbf{Privileged actors:} Access sensitive government datasets and develop targeted extraction capabilities. May deliberately include compromised identifiers or malicious associations in training data to enable tracking, surveillance, or discreditation of specific individuals or groups.
\end{itemize}

This adversary spectrum shows how identifier extraction and cross-context linkage are now possible even with minimal resources.

\section{Desired properties}
\label{sec:desired-prop}

To design the ideal properties for identifiers in light of our threat model, we first begin with a description of the characteristics of ULIs that make them particularly vulnerable to modern AI systems. These traits produce strong cross-context signals that can be exploited by downstream AI tooling:

\begin{itemize}[leftmargin=*]

    \item \textbf{Cross-context reuse}: The same ULI is used across multiple services, enabling cross-domain linkage. This creates a persistent join key that adversaries—especially breach aggregators and model trainers—can exploit to build comprehensive profiles across contexts, as modeled in our threat framework.

    \item \textbf{Recipient-side accumulation}: ULIs are observed repeatedly by recipients, enabling linkage across time and users. Regulatory mandates often require certain institutions to collect and retain nearly every ULI in circulation, concentrating exposure and making large-scale breaches likely. This accumulation directly amplifies the threat surface described in our model.

    \item \textbf{Temporal persistence}: ULIs persist for years or decades, increasing the likelihood that they are exposed in a breach and subsequently ingested into AI training pipelines or used for record linkage.
    
\end{itemize}

These characteristics arise from institutional design choices and legal mandates. As a result, large-scale breaches and data aggregation must be treated as baseline assumptions. Combined with the extraction, memorization, and inference capabilities of AI systems described in our threat model, identity systems must assume exposure and neutralize the resulting risks.

To address these vulnerabilities, we propose eight essential properties for privacy-preserving identifiers:

\begin{itemize}[leftmargin=*]

    \item \textbf{Forward unlinkability:} Compromised identifiers must not be linked to future interactions. This property is critical in the AI era where extracted identifiers may persist indefinitely in model weights, requiring that future interactions remain protected even after earlier exposures.
    
    \item \textbf{Per-relying-party unlinkability:} Each service receives a distinct, unlinkable identifier for the same individual. This prevents the cross-context joins that enable comprehensive profile building, even when multiple services' data is compromised or ingested into AI systems.
    
    \item \textbf{Relying party anonymity:} Issuers cannot track where or how identifiers are used, preventing surveillance. Identity assertions must be constructed and presented without issuer interaction during usage, as even metadata about service usage can be leveraged by AI for inference attacks.
    
    \item \textbf{Per-interaction unlinkability:} Interactions with the same service are unlinkable unless explicitly enabled. This helps prevent linking of interactions with the same entity if desired.
    
    \item \textbf{Easy delegation:} Support for user-controlled delegation (e.g., power of attorney, caregivers) without creating new linkages between identifiers or enabling correlation between the delegator and delegatee, which traditional systems frequently expose.
    
    \item \textbf{Minimal disclosure:} Proving specific attributes (e.g., "age > 18") without revealing complete credentials or usage context. This reduces the attack surface for AI memorization by limiting exposed data and prevents the issuer from tracking credential usage, addressing both service provider and issuer privacy concerns.
    
    \item \textbf{Verifiability:} Supporting cryptographic proofs of eligibility and compliance without compromising unlinkability. This allows for regulatory oversight and security enforcement while preserving the privacy guarantees needed to resist AI-based correlation.
    
    \item \textbf{Workflow compatibility:} Crucially, to achieve widespread adoption and displace vulnerable ULIs, the new system must offer pragmatic workflow compatibility~\cite{eid-poor-adoption}. This includes supporting existing practices such as entering alphanumeric identifiers into forms and documents.
\end{itemize}

Some identity systems partially address these issues with domain-specific pseudonyms~\cite{german-id}, and temporary identifiers~\cite{uidai_vid}. Anonymous credentials~\cite{camenisch2001efficient,alpar2017irma} theoretically achieve these privacy properties but disrupt established workflows. The AI era demands all properties simultaneously, as enhanced correlation capabilities can link even scoped identifiers to permanent ones~\cite{hansen2008privacy}. Our approach delivers comprehensive privacy while enabling drop-in replacement in existing workflows where identifiers need to be entered or printed in forms.

We next demonstrate a conceptual construction that achieves these properties while providing a straightforward migration path for systems currently using ULIs.

\section{Towards a cryptographic framework for unlinkable identifiers}

Having established the critical need for identifiers that are structurally resilient to AI-driven threats, we now outline the core principles and conceptual components of a cryptographic approach that could achieve the desired properties. The following is not intended as an exhaustive protocol specification, but rather as a demonstration of plausibility.

\subsection{System components}

\begin{itemize}[leftmargin=*]
    \item \textbf{Participant-generated identifiers and commitments:} Participants generate private sets of unlinkable identifiers and commit to them using a Merkle tree, with the root serving as their \textit{identity commitment}. The individual identifiers are utilized in various workflows where identifiers need to be presented and the participants have full control on the level of linkability they want to enable. For instance, they can present a new identifier per interaction, or utilize a scoped static identifier with a trusted party.

    \item \textbf{Coordinator-maintained lists:} Coordinators publish two Merkle roots:
    \begin{itemize}
        \item An \textit{allow root} derived from valid identity commitments
        \item A \textit{block root} from a Sparse Merkle Tree mapping revoked identity commitments to $\mathsf{true}$
    \end{itemize}

    \item \textbf{Zero-knowledge verification:} Participants present an identifier when demanded by a workflow and present a proof of legitimacy when needed by demonstrating in zero-knowledge that an identifier belongs to a valid, non-revoked identity commitment without revealing the commitment itself.

\end{itemize}

\subsection{Conceptual system operation}

Our framework is designed to serve as a drop-in replacement for ULIs while neutralizing AI-era threats. The operational flow is as follows:

\begin{enumerate}[leftmargin=*]
    \item \textbf{Participant action}: An individual generates a private, diverse portfolio of unlinkable identifiers, potentially for different contexts or individual interactions. They cryptographically commit to this entire portfolio via a Merkle tree, whose root becomes their \textit{identity commitment}. This initial step ensures self-custody of identifiers and prevents the formation of a singular, static ULI target that AI systems can easily track or memorize from breaches.

    \item \textbf{Coordinator action}: The participant's \textit{identity commitment} (not the identifiers themselves) is registered with a coordinator. This coordinator maintains public, auditable \textit{allow} and \textit{block} lists (represented by Merkle roots whose leaves are individual identity commitments). This phase enables necessary institutional oversight and revocation capabilities without granting the coordinator access to raw identifiers or fine-grained activity data that could be fed into AI analysis pipelines.

    \item \textbf{Privacy-preserving interaction and workflow integration}: When an identifier is required by a relying party, to be entered into a form, printed on a document for instance, the participant selects an appropriate one from their private portfolio. This selected identifier can be a standard alphanumeric string, appearing like a traditional ULI. Subsequently, or concurrently if the workflow demands immediate verification, the participant generates a zero-knowledge proof (ZKP). This ZKP attests that the chosen identifier is legitimate (i.e., part of their valid, non-revoked \textit{identity commitment}) \textit{without revealing the identity commitment itself} or any other identifiers in their portfolio. Relying parties verify this proof offline against the coordinator's allow and block roots. This ensures verifiability while severing the links between interactions. The ZKP can incorporate additional interaction-specific information.

\end{enumerate}

This operational model illustrates a path towards achieving the desired properties articulated in Section~\ref{sec:desired-prop} that eliminate the privacy issues caused by ULIs.

\section{Cryptographic sketch}

\subsection{Cryptographic primitives and notation}
We build our system using the following standard cryptographic primitives:

\begin{itemize}[leftmargin=*]
    \item \textbf{Cryptographic hash function} $\mathsf{CHF}: \{0,1\}^* \rightarrow \{0,1\}^l$ with standard collision and pre-image resistance properties~\cite{goldwasser2008lecture}.
    
    \item \textbf{Merkle tree}~\cite{merkle1980protocols} with leaves $l_0, \ldots, l_{m-1}$ and root $\mathcal{M} \leftarrow \mathsf{MerkleRoot}(\{l_i\})$. An inclusion proof $\mathsf{Proof}_{\mathcal{M}}(l_i)$ verifies leaf membership by reconstructing the root.
    
    \item \textbf{Sparse merkle tree (SMT)}~\cite{laurie2012revocation} mapping binary keys to values. Each key corresponds to a unique leaf; unused keys map to $\mathsf{false}$. A proof $\mathsf{Proof}_{\mathcal{R}}(k, v)$ demonstrates that $k \mapsto v$ under root $\mathcal{R}$. Non-inclusion of a key $k$ is proven by showing $\mathsf{Proof}_{\mathcal{R}}(k, \mathsf{false})$.
    
    \item \textbf{zk-SNARK}~\cite{goldwasser1985zk,groth2016snark,ggpr2013snark} enabling a prover to prove knowledge of witness $w$ satisfying relation $R(x, w)$ without revealing $w$. The prover generates $\pi \leftarrow \mathsf{Prove}(C, x, w)$, which the verifier checks using $\mathsf{Verify}(\pi, C, x) \rightarrow \{\mathsf{accept}, \mathsf{reject}\}$. Here, $C$ is the arithmetic circuit encoding the relation.

\end{itemize}

\subsection{Identifier generation and commitment}
Each participant generates a private set of unlinkable identifiers $\{id_0, \ldots, id_{n-1}\}$, either randomly~\cite{rfc4122} or deterministically from secrets~\cite{bip32,rfc_hkdf}. Deterministic generation provides a critical migration path: existing ULIs can be used as seed material to derive new unlinkable identifiers, enabling gradual transition from legacy systems. These methods support unlimited unique identifiers, eliminating reuse requirements. Identifiers may also encode commitments to attributes for selective disclosure.

The participant commits to these identifiers by constructing a Merkle tree: $\mathcal{I} \leftarrow \mathsf{MerkleRoot}(\{id_0, \ldots, id_{n-1}\})$. This identity commitment $\mathcal{I}$ is submitted to a coordinator for inclusion in the allow list. The coordinator never sees the underlying identifiers—only the root that anchors all future legitimacy proofs.

Optionally, the participant may bind the commitment to a persistent authenticator such as a public key or biometric measurement: $\mathcal{I}' \leftarrow \mathsf{CHF}(\mathcal{I} \,\|\, authenticator)$. This generic identity binding provides additional security while maintaining privacy. The separation ensures participants control identifier generation while coordinators merely publish allow and block roots.

\subsection{Authorization and revocation}
After constructing an identity commitment $\mathcal{I}$, a participant submits it to a coordinator for inclusion. The coordinator aggregates accepted commitments into a Merkle tree, publishing the root as the \emph{allow root}: $\mathcal{A} \leftarrow \mathsf{MerkleRoot}(\{\mathcal{I}_0, \ldots, \mathcal{I}_m\})$. This enables participants to prove authorization by presenting a Merkle inclusion proof $\mathsf{Proof}_{\mathcal{A}}(\mathcal{I})$ with respect to $\mathcal{A}$.

For revocation, the coordinator maintains a Sparse Merkle Tree mapping commitments to boolean flags. The root is published as the \emph{block root} $\mathcal{B}$. A commitment is revoked if it maps to $\mathsf{true}$; otherwise, it implicitly maps to $\mathsf{false}$. Participants prove non-revocation by presenting a proof $\mathsf{Proof}_{\mathcal{B}}(\mathcal{I}, \mathsf{false})$ with respect to $\mathcal{B}$.

These two roots, $\mathcal{A}$ and $\mathcal{B}$, jointly define the set of authorized, non-revoked commitments. They are periodically updated and publicly auditable. Coordinators never learn how commitments are used or which identifiers derive from them.

\subsection{Zero-knowledge proof of identifier legitimacy} 

To prove an identifier's legitimacy, a participant demonstrates it belongs to a valid, unrevoked identity commitment without revealing the commitment itself. Algorithm~\ref{alg:one-time-legitimate-id} defines our zero-knowledge circuit \textsc{VerifyIdentifier}. We exclude identity bindings and interaction-specific gadgets choosing to exclusively focus on identifier use.

The circuit proves that identifier $id$ belongs to an identity commitment that is both included in the allow list and absent from the block list. Critically, the commitment itself remains completely hidden from verifiers, preventing any cross-context linking. This circuit can optionally verify binding to an authenticator (e.g., public key or biometric tag) via a cryptographic hash, enabling features like biometric anchoring without compromising unlinkability.

To use a one-time identifier, the participant selects a pre-generated identifier, prepares the witness $w$ and public input $x$ as described, and computes $\pi_{id} \leftarrow \mathsf{Prove}(\textsc{VerifyIdentifier}, w, x)$. A verifier confirms legitimacy using $\mathsf{Verify}(\pi_{id}, \textsc{VerifyIdentifier}, x)$, which verifies that $id$ belongs to a hidden commitment in the allow list $x.\mathcal{A}$ and not in the block list $x.\mathcal{B}$, without ever revealing which commitment contains the identifier.

\begin{algorithm}[t]
    % \smaller
    \SetKw{Witness}{witness}
    \SetKw{PublicInput}{public-input}
    \SetKw{Assert}{assert}
    \textbf{Witness:}
    \begin{itemize}[noitemsep,nolistsep,leftmargin=*]
        \item $id$: identifier (Merkle leaf)
        \item $\mathcal{I}$: identity commitment
        \item $\mathsf{Proof}_{\mathcal{I}}(id)$: proof that $id \in \mathcal{I}$
        \item $\mathsf{Proof}_{\mathcal{A}}(\mathcal{I})$: proof that $\mathcal{I} \in \mathcal{A}$
        \item $\mathsf{Proof}_{\mathcal{B}}(\mathcal{I}, \mathsf{false})$: proof that $\mathcal{I} \notin \mathcal{B}$
    \end{itemize}
    
    \BlankLine
    \textbf{Constraints:}
    \begin{itemize}[noitemsep,nolistsep,leftmargin=*]
        \item Verify proofs: $\mathsf{Proof}_{\mathcal{I}}(id)$, $\mathsf{Proof}_{\mathcal{A}}(\mathcal{I})$, $\mathsf{Proof}_{\mathcal{B}}(\mathcal{I}, \mathsf{false})$
        \item Assert public inputs match witness: $id$ in $\mathsf{Proof}_{\mathcal{I}}(id)$, $\mathcal{A}$ in $\mathsf{Proof}_{\mathcal{A}}(\mathcal{I})$, $\mathcal{B}$ in $\mathsf{Proof}_{\mathcal{B}}(\mathcal{I}, \mathsf{false})$
    \end{itemize}
    
    \BlankLine
    \textbf{Public Input:}
    \begin{itemize}[noitemsep,nolistsep,leftmargin=*]
        \item $id$: presented identifier
        \item $\mathcal{A}$: allow root
        \item $\mathcal{B}$: block root
    \end{itemize}
    \caption{$\textsc{VerifyIdentifier}$}
    \label{alg:one-time-legitimate-id}
\end{algorithm}

\subsection{Properties satisfied}
The proposed system satisfies the properties defined in Section~\ref{sec:desired-prop}.

\begin{itemize}[leftmargin=*]
    \item \textbf{Unlinkability (forward, per-relying-party, per-interaction):} Achieved by generating distinct identifiers per recipient and interaction, each with separate proofs.
    
    \item \textbf{Privacy (relying party anonymity, minimal disclosure):} Ensured as proofs contain only the identifier and roots, with no recipient-specific information.
    
    \item \textbf{Delegation:} Enabled by Merkle subtree assignment, allowing surrogates to generate proofs independently.
    
    \item \textbf{Verifiability and workflow compatibility:} Maintained through efficient verification using public roots. The system supports standard identifier formats (e.g., numeric strings, UUIDs) that can be entered into existing forms and databases, enabling seamless integration with legacy systems while preserving the cryptographic security properties.
\end{itemize}

\begin{figure}[h]
\centering
\includegraphics[width=0.8\columnwidth]{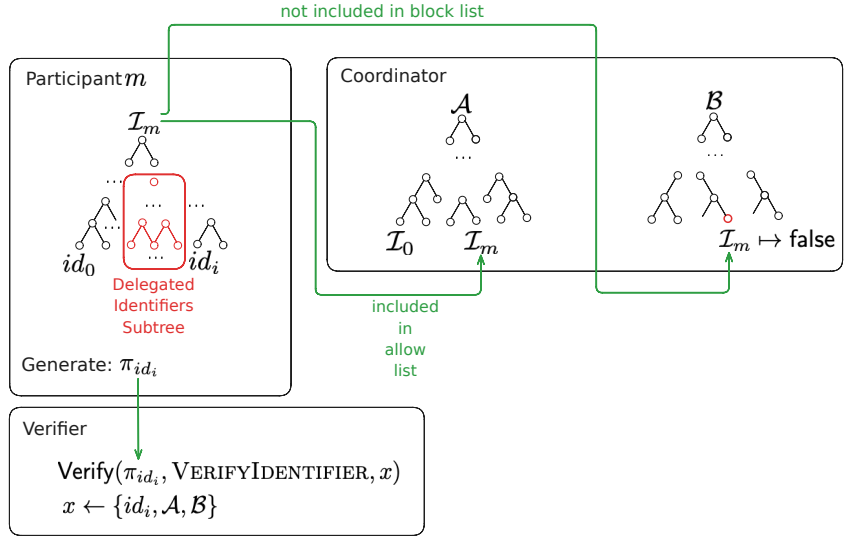}
\caption{Overview of our unlinkable identifier framework showing the three key components: participant-generated commitments, coordinator-maintained lists, and zero-knowledge verification.}
\label{fig:system_overview}
\end{figure}

Figure~\ref{fig:system_overview} illustrates how these properties are realized in our framework.

\section{Example workflows}
Our system enables both standard and previously impossible workflows in bureaucratic settings:

\begin{itemize}[leftmargin=*]
\item \textbf{Eligibility for state services:} Citizens prove eligibility for government services (e.g., benefits, voting) using distinct unlinkable identifiers per agency. Unlike today's ULI-based systems, this prevents cross-agency tracking while maintaining verifiable eligibility—the core function of identification.

\item \textbf{Anonymous medical consults:} Patients generate one-time identifiers for sensitive consultations without linkage to permanent records—impossible with current medical record systems that mandate persistent identifiers across all interactions.

\item \textbf{Regulatory compliance:} Participants prove they aren't on sanctions lists without revealing identity. This enables privacy-preserving compliance verification currently impossible under KYC/AML regulations~\cite{ofac2019framework} that require exposing full identities.
\end{itemize}

\section{Limitations and scope}
While our system offers significant privacy benefits, important limitations remain:

\begin{itemize}[leftmargin=*]
\item \textbf{Computational requirements:} Zero-knowledge proofs demand substantial client-side computation, making our approach incompatible with passive physical credentials (e.g., ID cards) and requiring appropriate hardware capabilities.

\item \textbf{Implementation complexity:} Coordinators must manage cryptographic commitments at scale, requiring infrastructure beyond current identity systems. Careful optimization of proof generation and verification is necessary for practical deployment.

\item \textbf{Protocol formalization:} Our approach requires explicit formalization of disclosure and audit policies as cryptographic protocols, increasing design complexity compared to ad-hoc methods in current systems.

\item \textbf{Secret management:} As with any cryptographic system, safe secret storage remains challenging for non-technical users.

\item \textbf{Incremental deployment:} The full privacy benefits emerge only when all services adopt the system. During transition periods, correlation between legacy and new identifiers remains possible.
\end{itemize}

These limitations constitute engineering and deployment challenges rather than fundamental barriers, suggesting directions for future work.

\section{Broader impacts}

Our work advocates the phasing out of ULIs and (i) introduction of unlinkable identifiers, (ii) zero knowledge proofs generated by participants to demonstrate the legitimacy of these identifiers. Besides directly addressing the threat model, we believe that anonymous identity systems and workflows like the ones proposed (i) restore individual agency - providing real choice for the privacy conscious, (ii) enhance trust in institutional interactions for individuals and marginalized communities especially, and (iii) reduce structural asymmetries in which individuals are permanently exposed, but institutions remain opaque~\cite{solove2006taxonomy}.

While such solutions enhance individual liberty and sovereignty, anonymity enhancing technologies have been used to cause harm~\cite{europol2020iocta} including distribution of CSAM and illicit financial activity. While we provide revocation mechanisms in our system, we acknowledge that the real world is more complicated than any list-based model allows. Entire classes of adversaries like foreign entities outside a jurisdiction might not ever have been enrolled in an identifier system but yet need to be blocked. Our intention with this position paper is not to present a complete regulatory apparatus, but to promote discussion about the future of identity in the AI era: (i) privacy risks are significantly amplified by modern AI capabilities, (ii) traditional safeguards such as redaction and access control are no longer sufficient, (iii) identity workflows must migrate toward cryptographic protocols that provide structural privacy, and (iv) such protocols are no longer theoretical or impractical.

\section{Conclusion}
\textbf{Universal Lifelong Identifiers (ULIs) are fundamentally unfit for the AI era and must be phased out in favor of unlinkable, cryptographically scoped identifiers.} This position paper demonstrates that modern AI capabilities have irreversibly broken the privacy assumptions underlying ULIs, creating unprecedented and uncontainable risks.

We have shown that comprehensive privacy is achievable without sacrificing functionality through a cryptographic architecture that supports verifiability, delegation, and regulatory compliance while preventing cross-context linking. Our approach offers a practical migration path from current systems, requiring minimal changes to established workflows.

Identity systems must evolve from privacy through policy to privacy through cryptography. In the AI era, where extraction capabilities are democratized and exposure becomes permanent, only structurally unlinkable identifiers can provide lasting protection. The framework presented here offers a viable path forward.

\bibliographystyle{plainnat}
\bibliography{references}

\end{document}